# Ferro-deformation and shape phase transitions over the nuclear chart: 50 ≤ protons (Z) ≤ 82 and 50 ≤ neutrons (N) ≤ 126

Chang-Bum Moon*

*Hoseo University, Chung-Nam 336-795, Korea*

April 3, 2016

We study a global nuclear structure in the framework of experimental observables. With the aid of large nuclear structure data at the national nuclear data center, NNDC, we present the distinctive systematic patterns emerged in the first $2^+$ excited energies, $E(2^+)$ and their energy ratios to the first $4^+$ levels, $R = E(4^+)/E(2^+)$, in the even-even nuclei, over $50 \leq Z \leq 82$ for protons, and $50 \leq N \leq 126$ for neutrons. We introduce the so-called *pseudo-shell* configurations from the subshells mixture in order to explain a semi-double shell closure, a shape phase transition, and a reinforced deformation. It is found that the reinforced deformation arises when Z = 64 or 66 correlates with N = 90 and reaches its maximum, indicating R = 3.3. Such a saturated reinforced deformation spans over Z = 58 to 72 and N = 100 to 106 as showing its center at Z = 64 or 66 and at N = 102 or 104. We define this reinforced deformation 'a *ferro-deformation*' like a ferro-magnetism in condensed matter physics. The shape coexistence would be expected to occur, such as a ferro-deformation, with a strong rotational mode, and a near spherical shape, with a vibrational mode, at the critical points of Z = 64 or 66, with N = 88 and 90; $^{150}$Sm and $^{152}$Sm, $^{152}$Gd and $^{154}$Gd, and $^{154}$Dy and $^{156}$Dy. We suggest that a super-deformation, which can be formed at high-lying excited states in a moderate deformed nucleus, would correspond to the ferro-deformation at N = 88 for the nuclei; Sm, Gd, and Dy. We argue that the ferro-deformation can be closely associated with a strong spin-orbital interaction between neutrons(ν) and protons(π) in the spin-orbit doublet, $\nu h_{9/2} - \pi h_{11/2}$, leading to the critical points at Z, N = 64, 104. The isospin dependent spin-orbital doublet interaction, having high-angular momentum, $l = 5$, drives nuclear surface toward a sudden and dramatic change, giving rise to a reinforced deformation. Of particular discussion is put on the nuclei at and close to Z = 50 for emphasizing the present systematic predictive power such that: $^{140}$Sn, with N = 90, should have a similar character to that of the adjacent isotopes; there is no sign revealing any phase transition in Ba between N = 88 and 90.

**Keywords**: national nuclear data center (NNDC), nuclear shell closure, pseudo-(sub)shell, shape phase transition, ferro-deformation, shape coexistence, isospin dependent spin-orbital interaction.

**Nuclides**: Ru, Pd, Cd, Sn, Te, Xe, Ba, Ce, Nd, Sm, Gd, Dy, Er, Yb, Hf, W, Os, Pt, Hg, Pb.

*cbmoon@hoseo.edu

## 1. Introduction

The shell structure of atomic nuclei is an outstanding example of shell structure patterns in the quantum world. The question of how shell structure develops in the finite quantum many-body systems has been a common problem among various disciplines; nuclear physics, atomic physics, condensed matter physics, molecular physics, and biophysics. Searching for fundamental characteristics of gross nuclear features based on large scale experimental observables is a good way toward basic understanding of the finite size many-body quantum mesoscopic system [1-6]. Especially systematic studies of the long isotones and isotopes sequences in terms of the first $2^+$ level energy, $E(2^+)$, provide some insights into a holistic understanding of the nuclear many-body systems. Figure 1 demonstrates such systematic features showing shell structure changes across the nucleon numbers. Moreover, we can see a close correlation between a pair of isotones, $_{52}$Te – $_{48}$Cd and $_{54}$Xe – $_{46}$Pd with respect to the proton closed shell $Z = 50$.

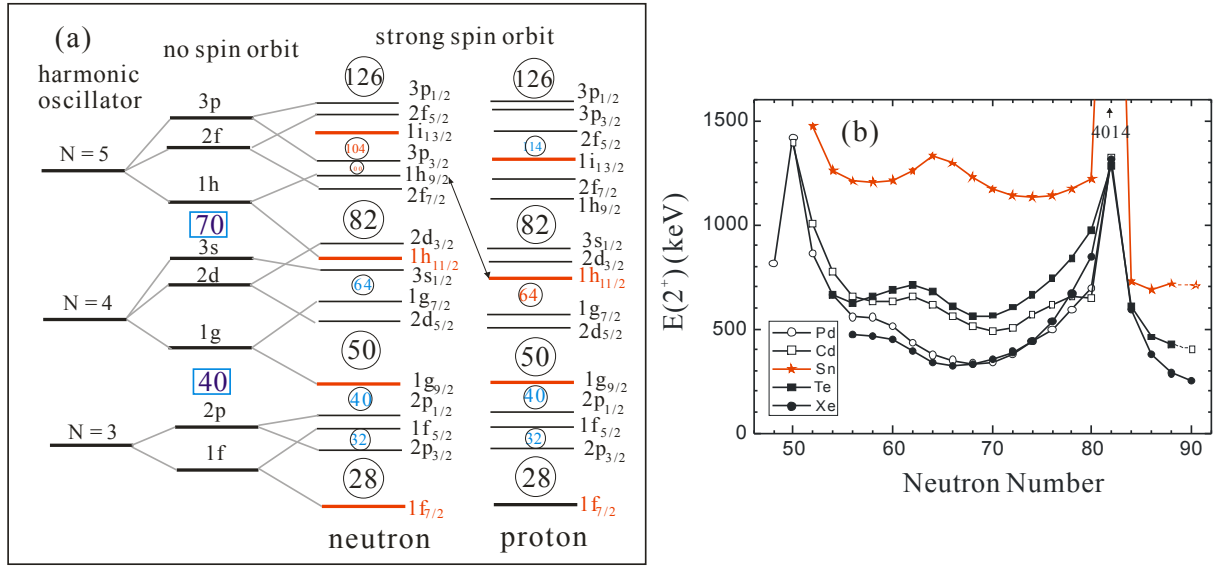

Fig. 1. (a) The single particle energies of a harmonic oscillator potential as a function of the oscillator quantum number N, a schematic representation of the single-particle energies of a Woods-Saxon potential, and a schematic illustration of the level splitting due to the spin-orbit coupling term. The numbers at the energy gaps are the subtotals of the number of particles represented by $N_j = 2j + 1$ of identical particles that can occupy each state. Note that the levels are numbered serially by a given orbital quantum number. For discussing, the associated isospin dependent spin-orbital interactions are denoted by the arrows and the corresponding critical points; $Z = 64$ and $N = 104$ or 100 are in red color. (b) Systematics for the first $2^+$ excited states in the nuclides in the vicinity of the shell closure $Z = 50$ (Sn) as a function of the number of neutrons. Data are primarily from [7], [8] for $^{126,128}$Pd, [9] for $^{136, 138}$Sn, [10] for $^{138}$Te, and [11] for $^{140}$Te. Points at N = 90, for Sn and Te, are the predicted values by the present work.

In this work, we examine the characteristics of nuclear shell structures at low-lying excited states based on the $2^+$ and $4^+$ level energies in the even-even nuclei. For practical purposes, we use the nuclear data at the national nuclear data center (NNDC) [7]. By extracting systematic behaviors of the first $2^+$ level energies, $E(2^+)$, and the ratio of the first $4^+$ level and $2^+$ level energies, $R = E(4^+)/E(2^+)$, we discuss the various nuclear structural phenomena over the nuclear chart $50 \leq Z \leq 82$ and $82 \leq N \leq 126$: pseudo-shell configurations, ferro(reinforced)-deformation, shape phase transitions, and shape coexistence. We point out the importance of isospin dependent spin-orbital interactions between neutrons in the $h_{9/2}$ orbital and protons in the $h_{11/2}$ orbital, for understanding both the nuclear shape phase transition and the ferro-deformation.

## 2. Ferro(reinforced)-deformation and shape coexistence

We begin by describing the systematic behavior of low-lying level properties in the even-even nuclei, as already shown in Fig. 1(b). A highly illuminating observable which adheres to systematic behavior is the excitation energy ratio of the first $2^+$ and $4^+$ states, $R = E(4^+)/E(2^+)$. This value, as a deformation parameter, evolves from < 2 for a spherical nucleus through 2 for a vibrator, to 3.3 for a deformed axial rotor [6, 12]. Figure 2 illustrates the systematics of R values for a given nucleus, along N = 48 to 90.

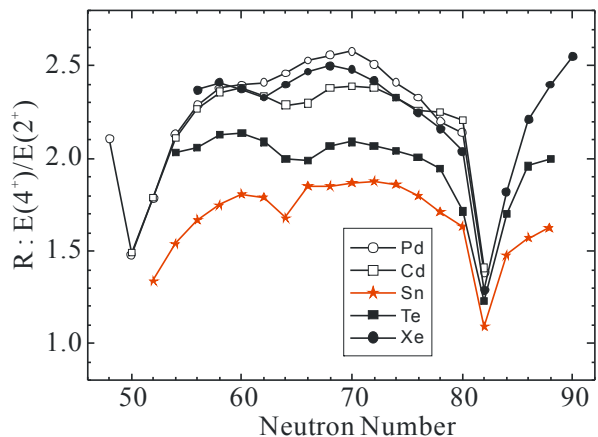

Fig. 2. Systematic plots of the R, $E(4^+)/E(2^+)$, values as a function of neutron numbers for the nuclei within Z = 46 and 54.

We notice first that for Sn, Te and Cd, two *pseudo-subshells* within N = 50 and N = 82 is formed as being separated at N = 64. Secondly, the R values of Cd are somewhat higher than those of Te, even though they are comparable in the $2^+$ level energies. The difference of R values between these two nuclides, with the same proton numbers, ± 2, with respect to Z = 50 shell closure, offers us some insights into understanding of how nuclear levels change according to the relative dominance between the particle-like nucleons and the hole-like nucleons. In view of the R value systematics, for the case of Cd as a soft deformed nucleus, both vibrational and rotational modes of collective excitation are expected. We argue that the Cd isotopes would be treated as a soft tri-axial rotator rather than a typical vibrator based on a spherical shape. For Xe and Pd, the two pseudo-shells become weaker as being merged around N = 70, resulting in one peak, with an intermediate deformation, R ≈ 2.5. The semi-subshell closure at N = 64 are due to a full-occupancy of a pseudo-subshell built on combining of the $d_{5/2}$ and $g_{7/2}$ subshells. Here we introduce a pseudo-subshell by denoting $J_{dg}(13/2)$, which indicates the configuration with a total spin of 13/2 in the mixed $d_{5/2}$ and $g_{9/7}$ subshells. In turn, the nuclides with 6 nucleons (holes) outside Z = 50, such as Ba and Ru are expected to have a more deformed shape with R > 2.5 as developing a peak at N = 64. This scenario is confirmed in Fig. 3, where one shell structure is evident as having its maximum value at N = 64. For the system of Ba-Ru, its deformation is a bit stronger in Ba than in Ru while for the Xe-Pd and for the Te-Cd, it is weaker in Z < 50 than in Z > 50. We point out that the very symmetry pattern with respect to N = 82 is unique in Te.

Now we extend our attention to Z = 82 and to N = 108 as shown in Fig. 4, in which the R values are plotted as a function of the number of protons. The emerging characteristics, firstly below N = 90, are as follows: First, for isotones above N = 82 a *pseudo subshell* within Z = 50 and 64 is formed; it is strong reinforced with N = 88, moderately with N = 86, and weakly with N = 84. It should be noted that this pseudo subshell corresponds to the pseudo-subshell we defined earlier, but in this case for protons, $J_{dg}(13/2)$ with capacitance of 14 protons occupancy. In contrast, there is no such a subshell responding to neutron numbers below N = 82. Second, the R values for Te are grouped into centered about two values, 2.0 (N = 76, 78, 86, 88) and 1.75 (N = 80, 84). This is quite a striking result that is not expected by the present theoretical, or empirical predictions. Moreover, an anomaly point at $^{132}$Te deviating the systematic trend along N = 80 line attracts our attention.

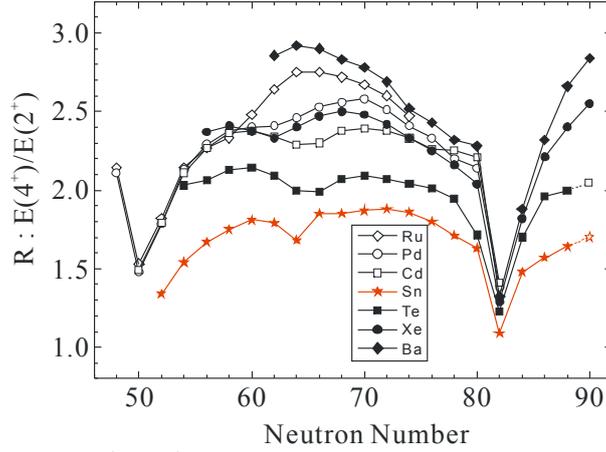

Fig. 3. Systematic plots for R (= E(4$^+$)/E(2$^+$)) values as a function of neutron numbers in isotopes between Z = 44 and 56. The points connected with dotted lines are an expected value following the systematics.

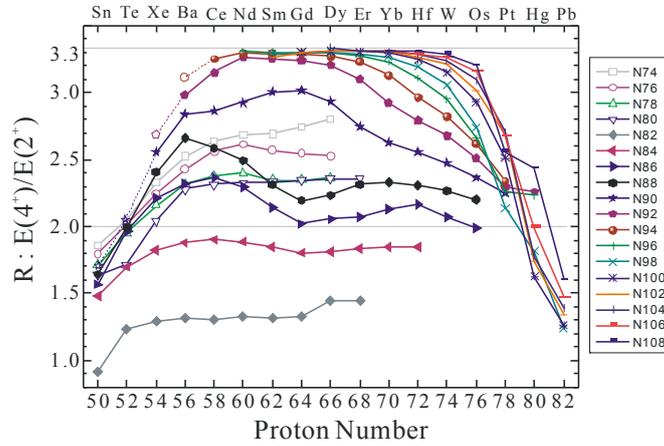

Fig. 4. Systematic plots for R (= E(4$^+$)/E(2$^+$)) values as a function of proton numbers in isotones between N = 74 and 100. The points connected with dotted lines are an expected value as obtained by the systematics. Data are primarily from [7], [13] for $^{164}$Sm and $^{166}$Gd, and [14] for $^{168}$Dy and $^{170}$Dy.

Next we focus on the characteristics at R(Z, N = 90). Surprisingly, the pseudo subshell developed at N = 88 does suddenly disappear at N = 90, revealing one shell feature with a peak around Z = 64. Such a sudden and dramatic change between N = 88 and 90 provides a clear signal for a shape phase transition, within the range of Z = 58 to 70. It is important to recognize that the region below Z = 56; for Te, Xe, and Ba, shows no structural changes even N ≥ 90. This implies that there is no structural phase transitions in Ba isotopes when N = 88 to 90. It is worthwhile to remember that the pseudo-subshell has the capacity of 14 protons (7 pairs of protons). Given that half-filled, high-j orbitals drive nuclear deformation, in the presence of a 50 ≤ Z ≤ 64 pseudo-subshell, Ba (Z = 56) or Ce (Z = 58) are expected to have maximal deformations. As shown in Fig. 4, this explains why $^{144}$Ba and $^{144}$Ce have a maximum value of R at N = 88 and at N = 86, respectively. Moreover, the second R maximum appears at Z = 70 for N = 88 and at Z = 72 for N = 86, respectively. This feature also explains existence of the second pseudo-subshell in the Z = 64 to 76 place. This pseudo-subshell mainly comes from a high-j angular momentum orbital, the h$_{11/2}$ subshell. A completion of this second pseudo-subshell, J$_{hds}$(17/2) requires three subshells; j$_{11/2}$, d$_{3/2}$, and s$_{1/2}$ with the capacitance of j = 17/2. From the present point of view based on the R parameter, we expect that no semi-double shell closure would appear to be at Z = 50 and N = 90, $^{140}$Sn.

We introduce another pseudo-subshell that is associated with the neutron orbitals for N > 82. The situation is very similar to the case of J$_{dg}$(13/2) since the f$_{7/2}$ and h$_{9/2}$ orbitals are readily combined so that they form a pseudo-subshell with j = 17/2. We denote this pseudo-subshell, as J$_{fh}$(17/2), with the capacitance of nucleons 18. Provided that half-filled at N = 90 or 92, it drives more strongly nuclear deformation as being reinforced with protons in the J$_{dg}$(13/2) pseudo-subshell. Under such high-deformed conditions, a *big pseudo-subshell* is formed by the combining of three $g_{7/2}$, $d_{5/2}$, and $h_{11/2}$ subshells. This large pseudo-shell has a capacitance of 26 protons having j = 25/2; $J_{dgh}$(25/2). Once this big pseudo shell is half- filled at Z = 64, the deformation reaches its maximum, R = 3.33 at N > 90.

The maximized R value between Z = 64 and 76, as shown in Fig. 4, extends more and more as increasing neutrons up to 104, which corresponds to the half-fill in the huge pseudo-shell within the N = 82 to 126 space. This strong reinforced deformation, finally, spans over 60 ≤ Z ≤ 74, centered at Z = 66 (or 64) with N = 104 (or 102). We define this reinforced deformation 'a ferro-deformation like a ferro-magnetism in condensed matter physics. This result implies that all the subshells between 50 ≤ Z ≤ 82 and 82 ≤ N ≤ 126 are involved coherently to build a big pseudo-shell with the capacitance of J = 31/2 and J = 43/2, respectively, giving rise to the ferro-deformation under both half-fill conditions.

We suggest that the shape coexistence would occur, such as a ferro-deformation, with a strong rotational mode, and a near spherical shape, with a vibrational mode, at the critical points of Z = 64 or 66 with N = 88 and 90; $^{150}$Sm and $^{152}$Sm, $^{152}$Gd and $^{154}$Gd, and $^{154}$Dy and $^{156}$Dy. We argue that the so-called super-deformation, which can be formed at high-lying excited states in a moderate deformed nucleus, would correspond to the ferro-deformation, indicating shape coexistence for Sm, Gd, and Dy, with N = 88.

We notice additionally that the region around Z = 78 and 80 shows an irregularity deviating a smooth systematic trend. This fluctuation may come from a competition of proton-neutron interactions between neutrons, with 100 < N < 106, favoring a deformed shape and protons, with 76 < Z < 86, favoring a spherical shape.

We raise a question why the ferro-deformation is formed over such a large area. An answer can be made by introducing the spin-orbital interactions between protons and neutrons in the spin-orbital doublet with the same orbital angular momentum, $l$ = 5; the $h_{9/2}$ for neutrons and the $h_{11/2}$ for protons, as producing the critical points at Z = 64 and N = 104. It should be noted that, according to the position of $p_{3/2}$ orbital as shown in Fig. 1, the neutron critical point swings between N = 104 and N = 100 as a sub-shell gap number. Such an isospin dependent spin-orbital interaction induces a sudden and dramatic shell structure change when N = 88 to 90, and gives rise to the ferro-deformation with a huge neutron pseudo-shell formation having the capacitance of J = 43/2. Our assumption is confirmed by the fact that there is no evidence indicating a pseudo-subshell formation below N = 82 as shown in Fig. 4. We summarize the results concerning the proposed pseudo-subshells and their characteristics in Table 1.

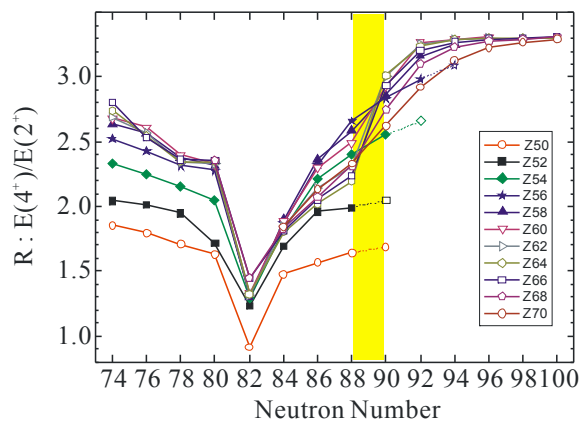

Fig. 5. Systematic plots for R (= $E(4^+)/E(2^+)$) values as a function of neutron numbers in isotopes between Z = 50 and 70. The points connected with dotted lines are an expected value in view of the systematics. The shaded region is in the occurrence of shape phase transitions. Notice also the very symmetrical pattern centered at N = 82 in the Te nuclei.

Now we again turn our attention to the R values as a function of neutrons, as shown in Fig. 5, in order to make a more wide visualization for the region beyond the shell closure N = 82. We also in this picture notice a distinctive characteristic feature in the region of N = 88 to 90; an abrupt increasing for Z = 60 to 68 as it results in the reverse of their deformation order between at N = 88 and at N = 90. This feature gives us again a signal for nuclear structural phase transitions at the critical points. In contrast, for Z = 50 to 56, Sn, Te, Te, Xe, and Ba, there is no sudden and dramatic change at all. With this point of view, we draw a reasonable conclusion that Ba has no phase transition from N = 88 to 90. The said phase transition in the literature would be relevant to the case of the nuclei along the down-hill side in the R curves at N = 88 or 86 in Fig. 4. For the case of Ba, the deformation parameter, R, increases gradually with no sign of change with increasing of neutron numbers, as reaching to about 3.0 at N = 94. Additionally, we suggest that the predicted R values are approximately 2.05 for Te at N = 90 and about 2.7 for Xe at N = 92, respectively. It should be remarked again with respect to the nuclei Te, as shown in Figs. 4 and 5, that the deformation parameter persists R ~ 2.0 at N = 88 or even, at N = 90.

Table 1. Summary of pseudo-shell configurations between double-shell closures of $50 \leq Z \leq 82$ and $82 \leq N \leq 126$: pseudo-shells, contributing subshells, total spins, half-filled nucleon numbers, and distinctive deformation configurations.

| pseudo-shells | combined subshells | pseudo-shell total spins: nucleon capacitance | half(full)-fill summed nucleon number | representative deformation configurations: (Z, N) |
|---|---|---|---|---|
| $J_{dg}(13/2)$ | $2d_{5/2}\ 1g_{7/2}$ | 13/2: 14 | 56 or 58 (64) | (56,88), (58,86) |
| $J_{hs}(13/2)$ | $1h_{11/2}\ 3s_{1/2}$ | 13/2: 14 | 70 or 72 (78) | (70,88), (72,86) |
| $J_{hd}(15/2)$ | $1h_{11/2}\ 2d_{3/2}$ | 15/2: 16 | 72 (80) | (72, 86) |
| $J_{hds}(17/2)$ | $1h_{11/2}\ 2d_{3/2}\ 3s_{1/2}$ | 17/2: 18 | 72 or 74 (82) | (72, 86), (74, 86) |
| $J_{dgh}(25/2)$ | $2d_{5/2}\ 1g_{7/2}\ 1h_{11/2}$ | 25/2: 26 | 62 or 64 (76) | (62, 92), (64, 90) |
| $J_{dghs}(27/2)$ | $2d_{5/2}\ 1g_{7/2}\ 1h_{11/2}\ 3s_{1/2}$ | 27/2: 28 | 64 (78) | (64, 102) |
| $J_{dghs}(31/2)$ | $2d_{5/2}\ 1g_{7/2}\ 1h_{11/2}\ 3s_{1/2}\ 2d_{3/2}$ | 31/2: 32 | 66 (82) | (66, 104) |
| $J_{fh}(17/2)$ | $2f_{7/2}\ 1h_{9/2}$ | 17/2: 18 | 90 or 92 (100) | (64, 90), (64, 92) |
| $J_{fhpi}(43/2)$ | $2f_{7/2}\ 1h_{9/2}\ 3p_{3/2}\ 1i_{13/2}\ 3p_{1/2}\ 2f_{5/2}$ | 43/2: 44 | 104 (126) | (64, 102), (66, 104) |

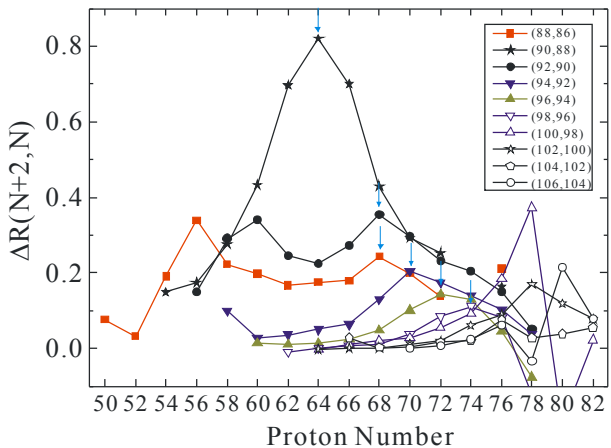

Fig. 6. The R values difference, ΔR(N+2, N) between two adjacent isotopes as a function of the number of protons. The arrows indicate peaks, indicating higher phase transitions, that move toward high Z with increasing N. Notice the peak, with one exception, of ΔR(92, 90) not at Z = 66 but at Z = 68.

Let us discuss in more details focusing on the features at the critical point, N = 90. In order to more visualize the presence of a phase transition, we deduce the R values derivative as shown in Fig. 6. Here the R value derivatives, ΔR(N+2, N) correspond to the R values difference between adjacent two isotopes with respect to neutrons. This parameter provides an amount of shape phase transitions between two adjacent isotopes. We can see a sharp peak at Z = 64 within Z = 56 and 70, signaling a sudden and dramatic change at ΔR(90, 88). How contrast it with that at ΔR(92, 90)! This behavior looks like a resonance phenomenon encountered frequently in the physical world. It is interesting to notice that the minimum value, close to zero, spans more and more with increasing the number of neutrons and finally spreads over 60 to 74 at ΔR(106, 104). A gradual shift of the peak starting at Z = 64 toward 74 is also noticeable, indicating a shape phase evolution in this region. Finally we can see large fluctuations for ΔR(100, 98) and ΔR(106, 104) in the region Z = 78 and 80.

On the basis of the above characteristics, we leave the following remarks: First, considering such a big difference of R values, shape coexistence is expected to be in Sm, Dy, and Gd with both N = 88 and 90. This is the same result as that obtained earlier. In view of such an irregularity at Z = 78 and 80, another shape coexistence is suggested to be in two Pt isotopes and two Hg isotopes, at N = 98 and 100 and N = 104 and 106, respectively. If we follow the present systematics, the points of ΔR(100-98) at Z = 78 and 80 should be around 0.1 and the points of ΔR(106-104) at Z = 98 and 80 should be near 0.05. An emergence of the second $0^+$, $2^+$, and $4^+$ levels at relatively low-lying excited states would confirm the very existence of two kinds of shape phase in the respective nuclei with N = 102, 104, and 106. More detailed discussions will be given in the forthcoming paper [15].

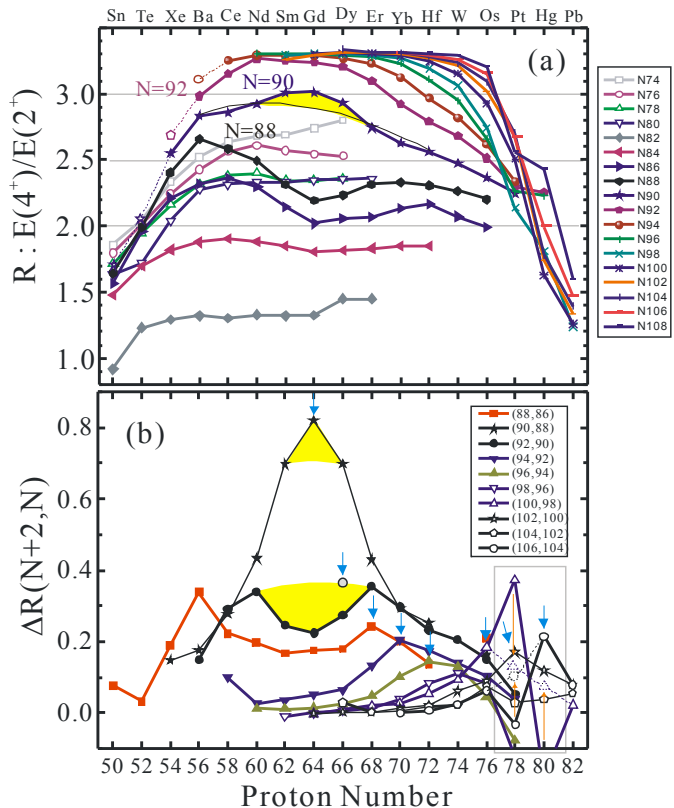

Fig. 7. The R and its derivative, ΔR(N+2, N) values at a given neutron number as a function of the number of protons. The shaded area corresponds to a difference between a presumed smooth variation (dotted line) and the experimental observables. A rectangular area indicates a possible shape coexistence region for the respective nuclei. See the text for more details.

For further discussion, we redraw Figs. 4 and 6, as shown in Fig. 7 where the shaded region is the difference between the predicted values, as obtained by a presumed smooth variation for N = 90, and the observables. If the smooth variation like a dotted line as shown in Fig. 7(a) is assumed, there would be neither a sharp peak nor a broad valley around Z = 64 as shown in Fig. 7(b). Consequently, occurrence of the peak and valley reflects a remarkable shape phase transition. Furthermore, if some deviating points in the region Z = 78 and 80 is rearranged for gaining a smooth variation, the peak at a given N, starting at Z = 64, moves gradually toward the higher proton number, to 80; see the arrows in Fig. 7(b). This feature also offers some insights about understanding of nuclear structure changes in terms of shape phase transitions and shape coexistence. The presence of the long-range correlations between protons and neutrons, especially in the same high angular momentum, and the competition of driving forces between them may be profoundly associated with the various changes of the low-lying excited states.

## 3. Structural characteristics of Sn, Te, Xe, and Ba beyond N = 82.

In this section, we study in more details on the nuclear structure properties for the nuclei in the vicinity of the shell closure Z = 50. Figure 8 demonstrates the R values difference, ΔR(82+N, 82–N) between a pair of isotopes with the same number of valence neutron holes and particles with respect to the closed shell N = 82. Accordingly, ΔR (88,76) means the R values difference between, as an example, $^{140}$Te with N = 88 and $^{128}$Te with N = 76. The emerging characteristics are as follows: for Te, there is little difference in the R values, giving ΔR ~ 0; for Sn, the differences are constant at ΔR ~ − 0.175; for Xe and Ba, the values are split into three regimes; positive, close to zero, and negative; for Ce ( Z = 58) to Dy ( Z = 66), the lines have two branches which extend into the negative region, and, finally, for N = 88 there are a distinctive peak at Z = 56 and a pronounced valley at Z = 64. This feature provides further confirmation of the existence of a pseudo-subshell between Z = 50 and Z = 64, strongly reinforced by N = 88. The above phenomenological arguments indicate, for Te there is no evidence of the associated nuclear structure change with deformation between N = 76 and N = 88. In other words, Te isotopes give a symmetrical signature that the same valence space results in a similar collectivity: *the valence neutron symmetry*. Moreover, an emergence of only negative values of ΔR indicates that Sn isotopes above N = 82 are less deformed than those below N = 82. The Ba isotopes show a greater deformation with N = 88 while $^{140}$Ba, with N = 84, is less deformed than their counter-part below N = 82, and $^{142}$Ba with N = 86 has the same in deformation as $^{134}$Ba with N = 78. The Xe isotopes follow similar systematics.

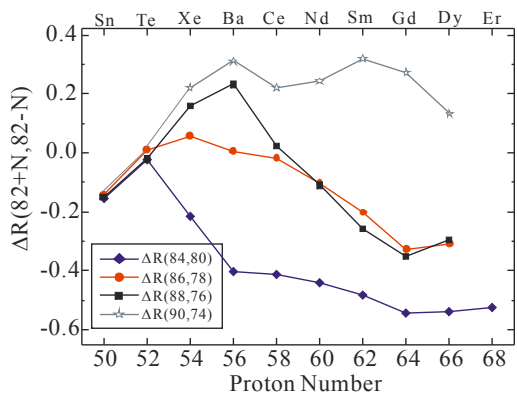

Fig. 8. Systematic plots for R (= E($4^+$)/E($2^+$) values difference, ΔR(82+N, 82–N; N = 0, 2, 4, 6) as a function of proton number in isotopes between N = 74 and 90. Here the R derivative, ΔR indicates correlations for a pair of isotopes with the same number of valence neutron holes and particles with respect to the closed shell N = 82.

If we extend our attention to N = 90, we can find that a sudden and dramatic change occurs in the vicinity of Z = 64 at ΔR(90, 74) and ΔR(88, 76). In addition, no difference between ΔR(88, 76) and ΔR(86, 78) illuminates a nuclear shape phase transition at ΔR(90, 88).

As a conclusion for the present work, we suggest the ground-state energy levels at N = 90 for Sn and Te as follows; as the $2^+ \to 0^+$ and the $4^+ \to 2^+$ transitions, (approximately) 700–490 keV for $^{140}$Sn and 400–420 keV for $^{142}$Te, respectively. See the corresponding figures; Fig. 1, Fig. 3, Fig. 4, Fig. 5, and Fig. 8. Also see Fig. 4(c) in Ref. 4.

Next we turn to systematic characteristics of the first excited $2^+$ states. As already pointed out, the systematics of $2^+$ states play an important role for investigating nuclear structure change from a singles spherical shape to a deformed collective shape. However, in order to obtain a quantitative aspect for the collectivity, it is better to draw the inverse value of $2^+$ energy, 1/E($2^+$). The reason is that the value of 1/E($2^+$) provides an estimation of the collectivity of a state on the basis of a proportionality of B(E2) to the value of 1/E($2^+$). This Grodzins rule [16], which relates small (large) transition energies to large (small) transition probability, works quite well in the nuclei around the stability line. Hereafter this value is denoted as a parameter by B(Z, N). We demonstrate the systematical trends of the B(Z, N) values and its derivatives, ΔB(82+N, 82–N) over 76 ≤ N ≤ 88 as a function of proton numbers in Fig. 9. Here the ΔB(82+N, 82–N) corresponds to the correlation parameter for the B values difference between a pair of isotopes with the same number of valence neutron particles and holes with respect to the N = 82 shell closure. Gross features for this systematics are very similar to those seen in the R and ΔR values: First, the systematics are divided in two regions, below and above N = 82. Second, the pseudo subshell also appears in the nuclei with N = 88 and 86 showing the maximum peak at Z = 56. Last but as a conclusion, the collectivity is stronger for 50 ≤ Z ≤ 68 in the region of N > 82 than that of N < 82. However, even the region N > 82, the measurements of B(E2) for $^{136}$Te [17] gave a negative result for this expectation since the measured B(E2) value was comparable to that at Z = 82. For gaining deeper insight into this contradictory, our next attention turns to the two neutron pairings and proton-neutron pairing interactions in large neutron-excess circumstances.

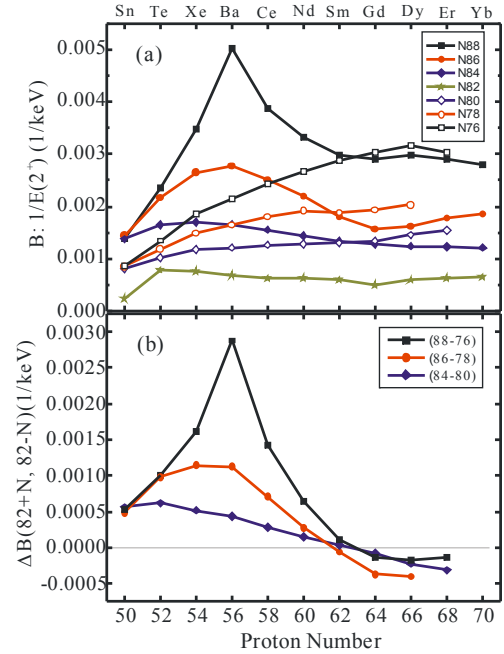

Fig. 9. Systematic plots for (a) B = $1/E(2^+)$ values as a function of proton number in isotopes between N = 76 and 88 and (b) the correlations for their differences between a pair of isotopes with the same number of valence neutron holes and particles with respect to the closed shell N = 82; ΔB(82+N, 82–N, N = 2, 4, 6).

The difference of the $2^+$ states energy between the isotopes below and above N = 82 provides an evidence for the different pairing force between protons and neutrons. Figure 10 describes the systematics based on such an energy difference, $\Delta E(2^+)$ between those with the same neutron holes below N = 82 and neutron particles above N = 82. We found, for example, the $\Delta E(2^+)$ values for a pair of isotopes for Sn versus Te nuclei between N = 80 and 84, N = 78 and 86, and 76 and 88 are 495 : 368, 481 : 379, and 426 : 320 in keV, respectively: demonstrating *the valence neutron asymmetry*. We summarize the results for Sn and Te isotopes in Table 2. As a result, the different pairing properties for Sn, Te, Xe, and Ba appear to be dependent on the ratios of the valence protons and the valence neutrons with respect to N = 82. Hence, the first $2^+$ states in neutron-rich nuclei above N = 82 are expected to have proton-neutron mixed asymmetric interactions as a pair of neutrons distributes dominantly over a pair of protons.

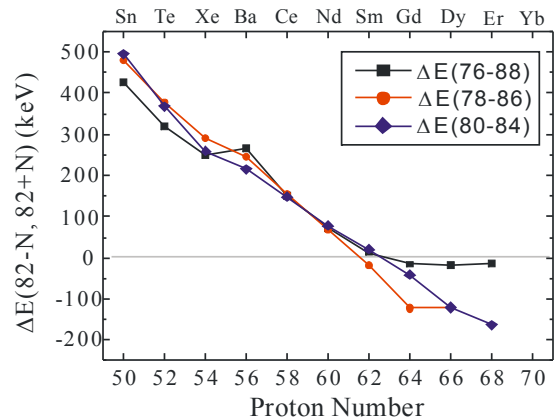

Fig. 10. Systematic plots for the $2^+$ states energy differences, ΔE(82–N, 82+N, N = 2, 4, 6) as a function of proton number between a pair of isotopes with the same number of valence neutron particles and holes with respect to the closed shell N = 82.

Table 2. The $2^+$ energy difference, $\Delta E(2^+)$ between a pair of isotopes for Sn and Te below and above N = 82. Energies are given in keV.

|  | 80 – 84* | 78 – 86* | 76 – 88* | Average |
|---|---|---|---|---|
| $\Delta E(2^+)$ for Sn | 495 | 481 | 426 | 467 |
| $\Delta E(2^+)$ for Te | 368 | 379 | 320 | 356 |
| $\Delta E(Te)/\Delta E(Sn)$ | 0.743 | 0.788 | 0.751 | 0.762 |

* Corresponding to the same neutron holes and particles with respect to N = 82.

Such a neutron-dominance in $2^+$ state leads to a weaker B(E2) value as already pointed out in $^{136}$Te [17]. Interestingly, the $\Delta E(2^+)$ values are almost identical within a variation of 10 % for both Sn and Te. This result implies that the neutron pairings at N = 86 and 88 are likely comparable to that at N = 84. From the $\Delta R$ values, we notice that Te isotopes maintain their collective character, showing a typical vibrator with $\Delta R = 0$ over the N = 82 region. We raise here the following questions; what of neutrons relative to a proton pair contribute, in energy, to the first $2^+$ state, and how does the neutron dominance in the $2^+$ state effect the B(E2) strength. A sophisticated shell model theory is required for understanding the underlying physics in $^{140}$Te, employing the proton-neutron mixed asymmetric interactions.

## 4. Conclusions

Based on the distinctive systematic features in the first $2^+$ excited energies, $E(2^+)$ and their energy ratios to the first $4^+$ levels, $R = E(4^+)/E(2^+)$, we demonstrated the emerging nuclear shell structure properties in the even-even nuclei, over 50 $\leq$ Z $\leq$ 82 for protons, and 50 $\leq$ N $\leq$ 126 for neutrons. By introducing the so-called *pseudo-shell* configurations from the subshells mixture, we explained various nuclear structure properties; a semi-double shell closure, a shape phase transition, and a reinforced deformation. The reinforced deformation starts when Z = 64 or 66 correlates with N = 90 and reaches R = 3.3, over Z = 58 to 70 and N = 96 to 100. Eventually, such a saturated reinforced deformation spans over Z = 58 to 72 and N = 100 to 106 as showing its center at Z = 64 or 66 and at N = 102 or 104. We defined this reinforced deformation 'a *ferro-deformation*' like a ferro-magnetism in condensed matter physics. This result implies that all the subshells between 50 $\leq$ Z $\leq$ 82 and 82 $\leq$ N $\leq$ 126 are involved in a coherent fashion to build a big pseudo-shell with the capacitance of J = 31/2 and J = 43/2, respectively, yielding the ferro-deformation under half-fill conditions. We suggest that shape coexistence would occur, such as a ferro-deformation, with a strong rotational mode, and a near spherical shape, with a vibrational mode, at the critical points of Z = 64 or 66 with N = 88 and 90; $^{150}$Sm and $^{152}$Sm, $^{152}$Gd and $^{154}$Gd, and $^{154}$Dy and $^{156}$Dy. We argued that the super-deformation, observed at high-lying excited states in a moderate deformed nucleus, would correspond to the ferro-deformation, which proves shape coexistence for Sm, Gd, and Dy, with N = 88. We suggested that the ferro-deformation would be closely associated with a strong spin-orbital interaction between neutrons in the $h_{9/2}$ orbital and protons in the $h_{11/2}$ orbital. This isospin dependent spin-orbital doublet interaction, with the high-angular momentum, $l = 5$, gives rise to suddenly a reinforced deformation, finally producing the ferro-deformation over the critical points at Z = 64 and N = 104. Moreover, such an isospin dependent spin-orbital interaction might play a critical role in constraints on the rapid neutron capture processes governed by Gamow-Teller beta-transitions. We point out some critical issues: there is no sign revealing any phase transition in Ba between N = 88 and 90; $^{140}$Sn, with N = 90, would have a similar character to that of the adjacent isotopes; and N = 70 should not contribute to forming a semi-double shell closure at Z = 40, $^{110}$Zr.

**Note added 1**: The suggestions and predictions concerning various nuclear structure aspects are made in the framework of the phenomenological arguments with the systematics of experimental observables distributed by the national nuclear data center, NNDC. Accordingly, many references, possibly related to the present work in the literature could not be quoted because of the referred data almost from NNDC.

**Note added 2**: Any abbreviation is avoided since it makes frequently the readers confusing like a jargon.